\documentclass[aps,prx,twocolumn,floatfix,showpacs,amsmath,amssymb]{revtex4-2}
\usepackage{graphicx}
\usepackage{hyperref}
\usepackage{xcolor}

\newcommand{\aver}[1]{\langle {#1} \rangle}
\newcommand{\es}[1]{\begin{split}#1\end{split}}
\newcommand{\bea}{\begin{align}}
\newcommand{\eea}{\end{align}}

\newcommand{\ii}{{\rm i}}
\newcommand{\ee}{{\rm e}}

\newcommand{\lp}{\left(}
\newcommand{\rp}{\right)}
\newcommand{\lsq}{\left[}
\newcommand{\rsq}{\right]}
\newcommand{\lbr}{\left\lbrace}
\newcommand{\rbr}{\right\rbrace}
\newcommand{\da}{\dagger}
\newcommand{\bma}{\begin{pmatrix}}
\newcommand{\ema}{\end{pmatrix}}

\newcommand{\mo}{{-1}}
\newcommand{\rw}{\rightarrow}
\newcommand{\oh}{\frac{1}{2}}
\newcommand{\w}{\omega}
\newcommand{\re}{\text{Re}}

\newcommand{\abs}[1]{ \left\lvert #1	\right\rvert}
\usepackage{bbold}
\newcommand{\id}{\mathbb{1}}
\newcommand{\hc}{\text{h.c.}}

\begin{document}

\title{How Subradiance Enables Nonlinearity in Weakly Driven Quantum Arrays}

\author{Orazio Scarlatella$^{1,2}$}
\author{Nigel R. Cooper$^{1}$}
\affiliation{$^{1}$TCM Group, Cavendish Laboratory, University of Cambridge, Cambridge CB3 0HE, United Kingdom\looseness=-1}
\affiliation{$^{2}$Sorbonne Université, CNRS, Institut des NanoSciences de Paris, 75005 Paris, France}


\begin{abstract}
Harnessing the nonlinear response of a medium is essential for applications including
frequency conversion and light amplification, as well as for the generation of quantum many-body correlations of light or matter. However, achieving these effects typically requires high
drive intensities and thick samples, which induce undesired heating effects that typically suppress
quantum correlations. In this work, we demonstrate that atom-thin arrays of quantum emitters
exhibit a robust nonlinear response even at arbitrarily weak drive intensities. This discovery
challenges the long-held assumption that weakly driven ensembles behave classically;
instead, we reveal that subradiant states provide a dominant nonlinear contribution that
persists in the low-intensity limit. Using a Dynamical Mean-Field Theory (DMFT) approach,
we predict that these nonlinearities generate a quantum-correlated steady state composed of
interacting pairs of subradiant excitations, characterized by long-range correlations and multi-mode squeezing. Our findings establish a new frontier for nonlinear quantum optics at
minimal power, and provide a scalable protocol for preparing multimode squeezing, offering
potential for applications in quantum metrology.
\end{abstract}
    \maketitle

Understanding the collective behaviour of an ensemble of quantum emitters coupled to the electromagnetic radiation is a long-standing problem in atomic physics and quantum optics \cite{dickeDicke1954,lehmbergLehmberg1970,agarwalAgarwal1970,grossHaroche1982}. 
When the inter-emitter distance is made comparable or smaller than the wavelength of the radiative transition, their dipolar interactions become strong and the emission collective, resulting in a strongly-interacting and dissipative quantum many-body problem. 
These subwavelength conditions lead to spectacular collective effects such as cooperative Lamb shifts \cite{friedbergManassah1973,keaveneyAdams2012}, superradiance and subradiance \cite{dickeDicke1954,grossHaroche1982}. 

The problem is receiving renewed attention, as different experimental platforms -- ultracold atoms \cite{srakaewZeiher2023,ruiBloch2020} and semiconductor dipolar excitons \cite{lagoinDubin2024,scuriPark2018,backImamoglu2018} -- are now entering the strongly-interacting regime.
Combined with the experimental ability to arrange quantum emitters in precise geometries,
the achievement of strong coupling results in an exquisite control of the light-matter interaction and of the emerging collective behaviour. These capabilities have recently led to remarkable applications, such as the realisation of single-layer mirrors of atoms \cite{bettlesAdams2016,shahmoonYelin2017,ruiBloch2020} -- with coherent and spatial control of their reflectivity \cite{srakaewZeiher2023} -- as well as of excitons \cite{zeytinogluImamoglu2017,backImamoglu2018,zhouPark2017,scuriPark2018}, contributing towards the realisation of metasurfaces that can control the quantum properties of light  \cite{bekensteinLukin2020,srakaewZeiher2023,solntsevKivshar2021}.

Applications, as well as most theoretical understanding, have been so far limited to the regime of linear response. Harnessing the nonlinear properties of quantum emitters arrays offers the potential for far-reaching consequences for quantum nonlinear optics~\cite{krasnokAlu2018,liZentgraf2017,solntsevKivshar2021,wildLukin2018}, as well as for quantum metrology \cite{henrietAlbrecht2019}, and quantum computing (where atom arrays are one of the most promising architectures \cite{manetschEndres2025,bluvsteinLukin2022,bluvsteinLukin2024}).

The subradiant eigenstates~\cite{asenjo-garciaChang2017,zhangMolmer2019,rubies-bigordaYelin2023} of these systems --
interesting as quantum memories \cite{plankensteinerGenes2015,facchinettiRuostekoski2016,asenjo-garciaChang2017,needhamOlmos2019,ferioliBrowaeys2021,ballantineRuostekoski2021,rubies-bigordaYelin2022,cechOlmos2023} and for quantum metrology \cite{ostermannGenes2013,facchinettiRuostekoski2018,quRey2019} -- provide an attractive basis where to probe strong nonlinear effects~\cite{asenjo-garciaChang2017,henrietAlbrecht2019,rubies-bigordaYelin2023}, owing to their heavily suppressed decay rates.
Nevertheless, these are difficult to excite and manipulate. In particular, they cannot be excited with a weak, (linearly-) resonant electromagnetic driving field~\cite{asenjo-garciaChang2017,heChang2020,zannerKirchmair2022} (they are dark states). 
This fact prevents the possibility of driving them weakly into nonlinear regimes, without generating substantial heating that typically largely suppresses quantum correlations.

The view that subradiant state are negligible has had the far-reaching consequence that the weak-drive regime of emitter ensembles has long been regarded as one of linear response, accurately described by classical equations of motion \cite{ruostekoskiJavanainen1997,javanainenJavanainen1999,leeRuostekoski2016,williamsonRuostekoski2020,bettlesAdams2015,bettlesAdams2016,parmeeCooper2018,sutherlandRobicheaux2016,robicheauxSuresh2021,glicensteinBrowaeys2020,guerinKaiser2016}, which have been viewed as an exact description of the limit of weak drive intensity~\cite{ruostekoskiJavanainen1997,javanainenJavanainen1999,leeRuostekoski2016,williamsonRuostekoski2020}.

In this work, we show that, by contrast, the weak-drive regime of a large array of quantum emitters is non-perturbative and surprisingly nonlinear. {This is due to resonant ``parametric driving'' processes, involving pairs of single-particle subradiant modes.
We first illustrate their effect by treating them perturbatively with respect to noninteracting theories~\cite{ruostekoskiJavanainen1997,javanainenJavanainen1999,leeRuostekoski2016,williamsonRuostekoski2020,bettlesAdams2015,bettlesAdams2016,parmeeCooper2018,sutherlandRobicheaux2016,robicheauxSuresh2021,glicensteinBrowaeys2020,guerinKaiser2016}.  
We show that this amounts to constructing a perturbation theory around a model of a parametric amplifier~\cite{scullyZubairy1997}, which becomes dynamically unstable beyond a threshold driving strength thereby signalling a non-perturbative steady state.
Because this threshold decreases to zero with increasing emitter number, the established classical theories ~\cite{ruostekoskiJavanainen1997,javanainenJavanainen1999,leeRuostekoski2016,williamsonRuostekoski2020,bettlesAdams2015,bettlesAdams2016,parmeeCooper2018,sutherlandRobicheaux2016,robicheauxSuresh2021,glicensteinBrowaeys2020,guerinKaiser2016} prove inadequate to describe the weak-drive behaviour of large arrays.

Then, we study the resulting steady-state under continuous driving, using the non-perturbative Dynamical Mean-Field Theory (DMFT) for Markovian spin systems recently introduced \cite{scarlatellaCooper2024,scarlatellaSchiro2021}. This has been so far applied only to the regime of strong drive intensities~\cite{scarlatellaCooper2024} -- in which the steady-state is a trivial infinite temperature state, albeit with an interesting fluorescence spectrum \cite{scarlatellaCooper2024}: here we address the more challenging~~\cite{minkFleischhauer2023,robicheauxSuresh2021} weak-driving regime, which required several technical improvements (discussed later in the main text, as well as in SM Notes \ref{sm:numerical_schemes} and \ref{sm:dmft_linear_stability}).

Using DMFT, we find that the weak-drive nonlinearities lead to a steady state of interacting pairs of subradiant excitations. 
This is characterised by quantum correlations with multimode squeezing and long-range character, which are robust to the heating caused by driving, dissipation and interactions. 
Moreover, we demonstrate that these nonlinear effects are strong: the nonlinear steady-state population is comparable with the linear component predicted classically. This confirms the inadequacy of classical theories~\cite{ruostekoskiJavanainen1997,javanainenJavanainen1999,leeRuostekoski2016,williamsonRuostekoski2020,bettlesAdams2015,bettlesAdams2016,parmeeCooper2018,sutherlandRobicheaux2016,robicheauxSuresh2021,glicensteinBrowaeys2020,guerinKaiser2016}, and indicates a high conversion efficiency of drive photons into the nonlinearly-populated modes, relevant for nonlinear optics.

Our findings can be realised in current experiments
with dipolar excitons \cite{lagoinDubin2024}, and in upcoming ones with
ultracold atoms \cite{ruiBloch2020,srakaewZeiher2023}, which will soon operate in suitable regimes. 
They open a new frontier for nonlinear quantum optics at minimal power and with atom-thin media, avoiding the significant heating generated in typical strong-driving regimes, thereby allowing a much larger degree of quantum correlations and quantum non-linear effects;
for example, we discuss how our setup could enable enhanced generation of entangled photons. 
Our work also provides a simple scheme to generate multimode squeezing correlations, offering potential for applications in quantum metrology, where such schemes and correlations are highly sought after~\cite{maNori2011}.

\begin{figure}
\centering    \includegraphics[width=1\linewidth]{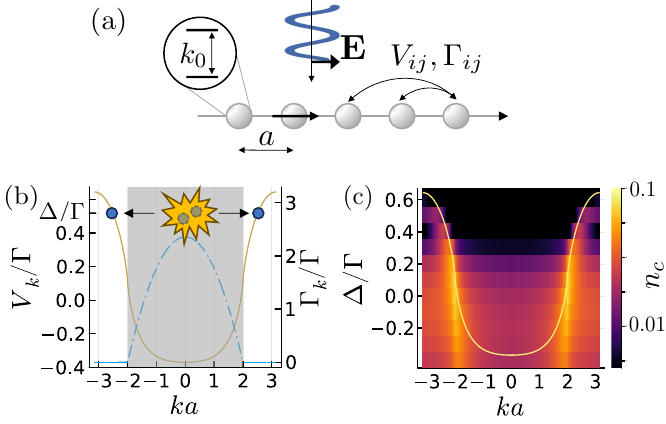}
\caption{(a) A schematic of a 1D array of two-level emitters, with transition wavevector $k_0$ and lattice spacing $a$, under external driving. The driving electric field vector $\textbf{E}$ is oriented along the chain, inducing the orientation of the atomic dipoles indicated by an arrow. The emitters are subject to coherent dipole-dipole interactions $V_{ij}$ and collective decay processes with rates $\Gamma_{ij}$, both of which are long-ranged. 
(b) The dispersion relation $V_k$ (solid line) and decay rates $\Gamma_k$ (dash-dot line) of collective single-particle modes for $N\rw \infty$ emitters, as functions of momentum $k$, for the subwavelength lattice spacing $k_0 a =2$: the modes for $\abs{k} > k_0 $ -- in the non-grayed area -- correspond to subradiant modes with zero decay rates (in the thermodynamic limit), that are linearly decoupled from the far-field; the illustration instead depicts a nonlinear process, in which two drive photons scatter resonantly into a pair of subradiant modes. 
(c) The number of steady state excitations $n_c(k) = \aver{\sigma_k^+ \sigma_k^-}_c$ calculated in DMFT/NCA for $N\rw \infty$, as a function of momentum and drive detuning $\Delta$ (the dispersion relations is superimposed in yellow). Here the drive strength is $\Omega/\Gamma = 0.8$, the lattice spacing is the same as in (b). It shows that the subradiant modes are resonantly populated. 
}
    \label{fig:schematics} \label{fig:fig1}
\end{figure}

\textit{Model.} We consider $N$ two-level emitters ordered in a one-dimensional (1D) geometry, and specifically the one illustrated in Fig. \ref{fig:fig1} (a). Most of our conclusions and methods can nevertheless be extended to more general 1D geometries, as well as to two dimensions. The emitters are illuminated with a uniform plane wave, and are coupled to the free-space electromagnetic vacuum, giving rise both to coherent dipole-dipole interactions and to collective decay.

Such system can be described by the Markovian master equation \cite{lehmbergLehmberg1970} 
\begin{align}
\label{eq:atLightME}
\dot{\rho} &=-\frac{\ii}{\hbar}[H, \rho]+\mathcal{D}[\rho] \\
\label{eq:ham_0}
H &=  -\frac{\hbar \Delta}{2} \sum_i^N {\sigma}_i^z + \frac{\hbar \Omega}{2} \sum_i^N {\sigma}_i^x + \sum_{i \neq j}^N  \hbar V_{i j} \sigma_i^{+} \sigma_j^{-}, \\  
\mathcal{D}[\rho]&=\frac{1}{2} \sum_{i, j} \Gamma_{i j}\left(2 \sigma_i^{-} \rho \sigma_j^{+}-\sigma_i^{+} \sigma_j^{-} \rho-\rho \sigma_i^{+} \sigma_j^{-}\right) . \label{eq:dissipat}
\end{align}
Here $\sigma_l^\alpha$ with $\alpha=x,y$ or $z$ are the Pauli matrices on site $l$, and ${\sigma}_l^{ \pm}={\sigma}_l^x \pm \ii {\sigma}_l^y$ the raising and lowering operators. $\Delta = \omega_d - \omega_0$ is the detuning of the driving-field frequency from the two-level transition energy $\omega_0$, and $\Omega=2 \mathbf{d} \cdot \mathbf{E} / \hbar$ is the Rabi frequency -- indicating the strength of the driving --, given by the vector of transition dipole moments $\mathbf{d}$ and by the driving electric-field vector $\mathbf{E}$. 
Such field is assumed uniform and oriented along the array (see Fig.~\ref{fig:fig1} (a)), leading to the rates of coherent dipolar interactions and collective decay
\begin{align}
\label{eq:vij}
V_{i l} & =-\frac{3 \Gamma}{2}\left[\frac{\sin k_0 r_{i l}}{\left(k_0 r_{i l}\right)^2}+\frac{\cos k_0 r_{i l}}{\left(k_0 r_{i l}\right)^3}\right], \\
\label{eq:gamij}
\Gamma_{i l} & =3 \Gamma\left[-\frac{\cos k_0 r_{i l}}{\left(k_0 r_{i l}\right)^2}+\frac{\sin k_0 r_{i l}}{\left(k_0 r_{i l}\right)^3}\right].
\end{align}
A key parameter determining these couplings is $k_0 a = 2\pi a/{\lambda_0}$, namely the ratio between the lattice spacing $a$ and the emitter transition wavelength $\lambda_0$ -- where $k_0$ is the transition wavevector. 
It controls the relative amplitude and range of the couplings, such that for decreasing $k_0a$ they become longer-ranged and the coherent interactions increase in magnitude, while for increasing values the regime of independent emitters is approached.
Here we consider intermediate values, for which both couplings are of the same order, and thus interesting driven-dissipative dynamics is expected. 
Hereafter, we assume units of $\Gamma = 1$ and $\hbar=1$.

\textit{Weak-drive and subwavelength regimes. } 
Eqs.~\eqref{eq:atLightME}-\eqref{eq:gamij} describe a many-body problem with long-range interactions and dissipation, which is brought out of equilibrium by external driving, and is in general hard to solve.

In the particular case in which a single excitation is prepared in the system, and in absence of driving $\Omega=0$, such equations become equivalent to a problem of noninteracting bosons. This is obtained by replacing the spin operators with bosonic ones $\sigma^-_j \rw a_j$ \cite{porrasCirac2008,asenjo-garciaChang2017,svidzinskyScully2010}, via a Holstein-Primakoff representation of the spin. It can be solved exactly in terms of eigenmodes of collective excitations, with frequencies shifted from those of individual emitters of $V_\alpha$, and with decay rates $\Gamma_\alpha$, where $\alpha$ is the mode index. 
For a non-zero yet weak driving, one may also approximate~\eqref{eq:atLightME}-\eqref{eq:dissipat} with such single-particle bosons theory -- upon adding a linear drive term --, provided that the individual population of the collective modes remains weak.
For a resonant monocromatic drive, this typically holds for a drive strength smaller than the driven mode decay rate $\Omega \ll \Gamma_\alpha$.

Additionally, in the subwavelength regime in which the emitters spacing is smaller than their transition wavelength, modes with decay rates much smaller than independent emitters $\Gamma_\alpha \ll \Gamma $ appear, which are called subradiant \cite{grossHaroche1982,asenjo-garciaChang2017}. These arise due to destructive interference of the field radiated by different emitters. 
Despite the long lifetimes, by the same interference mechanism, subradiant modes are also linearly decoupled from far-field drives, thus they are not excited under a weak driving field~\cite{asenjo-garciaChang2017,heChang2020,zannerKirchmair2022}.
We remark that this discussion is not specific to the case of arrays, but also extends to disordered ensembles, often realised in cold-atoms experiments \cite{ciprisBachelard2021, dasYavuz2020, ferioliBrowaeys2021, glicensteinFerrier-Barbut2022, guerinKaiser2016} -- also described by the model \eqref{eq:atLightME}-\eqref{eq:dissipat} upon specifying the appropriate couplings $V_{ij}$ and $\Gamma_{ij}$.

For these reasons, even in the subwavelength regime and in the presence of -- in principle, highly susceptible -- subradiant states, the weak-drive regime has been widely regarded as a linear regime that is accurately described by classical equations of motion \cite{ruostekoskiJavanainen1997,javanainenJavanainen1999,leeRuostekoski2016,williamsonRuostekoski2020,bettlesAdams2015,bettlesAdams2016,parmeeCooper2018,sutherlandRobicheaux2016,robicheauxSuresh2021,glicensteinBrowaeys2020,guerinKaiser2016}, equivalent to the noninteracting bosons description. These were expected to become exact descriptions in the limit of vanishing drive intensity \cite{ruostekoskiJavanainen1997,javanainenJavanainen1999,leeRuostekoski2016,williamsonRuostekoski2020}, without any distinction being made between the cases of disordered and ordered ensembles. This conclusion turns out to be wrong in the case of arrays of large sizes.

In this case, discrete translational invariance implies that lattice momentum $\textbf{k}$ is a good quantum number, and the bosonic eigenmodes correspond to reciprocal-space modes $a_\textbf{k} =  \sum_{ j } \ee^{\ii \textbf{k} \cdot\textbf{ r}_j} a_j /\sqrt{N}$, with dispersion relation and decay rates given by the Fourier transforms of  \eqref{eq:vij}  and \eqref{eq:gamij} $V_{\textbf{k}}$ and $\Gamma_{\textbf{k}}$. These are shown in Fig.~\ref{fig:fig1}~(b) for our geometry (for $N\rw \infty$). 
In the subwavelength regime, an entire manifold of subradiant eigenmodes arises, with decay rates strongly suppressed by the number of emitters. 
These decay rates have been found~\cite{asenjo-garciaChang2017} to depend polynomially on $N$ for open boundary systems $\Gamma_\textbf{k} \sim N^{-\alpha}$, with $\alpha$ seemingly universal for the most subradiant modes with $\alpha =3$ in 1D, and $\alpha = 6$ for a 2D square lattice. It was also shown that the residual decay is mainly due to scattering off the array boundaries, and that for periodic boundary conditions in 1D the suppression becomes exponential $\Gamma_\textbf{k} \sim \exp(-N)$. 

This suppression can be understood by subradiant modes having an energy much smaller than free-space photons with the same longitudinal momentum -- the momentum along the array --, thus their excitations not to be able to decay into single photons conserving energy and momentum \cite{asenjo-garciaChang2017,heChang2020}. 
From this argument, it follows that this happens whenever the mode dispersion intersects the light-cone within the Brillouin zone \cite{asenjo-garciaChang2017}, corresponding to the condition $k_0 a \leq \pi$ in 1D and $k_0 a \leq \sqrt{2} \pi$ for a 2D square lattice, and that those modes appear at large momenta $\abs{\textbf{k}}>k_0$, as in the case shown in Fig.~\ref{fig:fig1}~(b).

\textit{Perturbative considerations.} While the direct coupling of a far-field drive with subradiant modes is suppressed, resonant and efficient excitation of these modes can still happen through higher-order processes in perturbation theory, mediated by the nonlinearities of the atoms. 

These processes can be identified by expanding a Holstein-Primakoff representation of the spin in terms of bosons~(see SM Note \ref{sm:correlation_functions_relaxation}), under the assumption that the drive is sufficiently weak, such that the steady state has a low density of excitations around the state with all spins pointing down. 
One can then expand simultaneously in 
the drive and in the nonlinear terms, and retain only resonant (i.e. energy-conserving) processes -- as non-resonant ones are energetically suppressed. For our geometry -- with a drive direction orthogonal to the array --, the lowest-order such processes involve the pair of subradiant modes with momenta $k$ and $-k$, such that $V_{\pm k} = \Delta$. These read
$ (\Omega/\sqrt{N} a^\dagger_{-k}a^\dagger_{k} a_{k=0})_{t_2}  ({\sqrt{N} \Omega} a^\dagger_{k=0})_{t_1} $ and $ ({V_k}{/ N} a^\dagger_{-k}a^\dagger_{k} a_{k=0}a_{k=0})_{t_3}  ({\sqrt{N} \Omega} a^\dagger_{k=0})_{t_2} ({\sqrt{N} \Omega} a^\dagger_{k=0})_{t_1} $, where $a$ is the boson annihilation operator.
These processes involve an intermediate (virtual) excitation of the $k=0$ mode, and convert two drive photons (as by the $\Omega^2$ factors) into a pair of subradiant excitations, conserving energy and momentum (see the illustration in panel (b) of Fig. \ref{fig:fig1}). 
Importantly, lattice-momentum conservation ensures that the matrix element of these processes (between e.g. an initial vacuum state and the subradiant eigenstates) can also be non-negligible -- a consideration that is distinct from energetics. 
We note that this would not be the case in the case of disordered ensembles, to which we therefore expect our conclusions not to apply -- on the contrary we expect a similar phenomenology in arrays of other geometries.

To estimate the effect of the above processes, we first treat them perturbatively with respect to the noninteracting theory -- the single-particle bosonic model with an added coherent drive, equivalent to the classical approaches of \cite{ruostekoskiJavanainen1997,javanainenJavanainen1999,leeRuostekoski2016,williamsonRuostekoski2020,bettlesAdams2015,bettlesAdams2016,parmeeCooper2018,sutherlandRobicheaux2016,robicheauxSuresh2021,glicensteinBrowaeys2020,guerinKaiser2016} --, assuming a weak driving $\Omega$ and a large number of emitters $N$ (see SM Note \ref{sm:perturbative_calculations}).
In this respect, the fluctuations of the $k=0$ mode — which is directly, although off-resonantly driven — around its average value can be safely neglected. Keeping only the resonant pair of subradiant modes  (on the same basis as earlier), 
this leads to an effective model for them. 
The noninteracting part of such model has Hamiltonian $H_{k} = \delta_{k} (a_{k}^\da a_{k} + a_{-k}^\da a_{-k}) 
+ {\lambda}_{k} ( a_{k}^\da a_{-k}^\da +  \hc )$, where coefficients $\delta_k ,\lambda_k$ are independent of $N$ and of leading order $\Omega^2$ in the drive strength. 
The dissipator, at leading order in $\Omega$ and $N$, is simply obtained by replacing  $\sigma^- \rw a$ in \eqref{eq:dissipat} and is diagonal in momentum, where further corrections are negligible since the decay rates of subradiant modes are already suppressed for large $N$.

The noninteracting part of the effective model obtained is then a model of a parametric amplifier~\cite{scullyZubairy1997}. 
In addition, in our problem the modes involved -- owing to their subradiant nature -- have decay rates $\Gamma_{\pm k}$ that decrease to zero with $N$.
It is a well-known fact that a parametric amplifier develops a parametric instability for $\Omega \gtrsim \Gamma_{\pm k}^{1/2}$~\cite{scullyZubairy1997}, entailing a divergence of its steady-state moments. 
Then, in the case of our problem and for
the most subradiant modes, this instability takes place for $\Omega \gtrsim N^{-\alpha/2}$ for open boundary conditions, and $\Omega \gtrsim \exp(-N/2)$ for periodic ones.
Beyond such thresholds, the noninteracting model is dynamically unstable, and it therefore cannot be used as the basis of a perturbative analysis. 
This reveals that the steady-state of the full interacting problem~\eqref{eq:atLightME} is non-perturbative, and that the widely-used noninteracting theories~\cite{ruostekoskiJavanainen1997,javanainenJavanainen1999,leeRuostekoski2016,williamsonRuostekoski2020,bettlesAdams2015,bettlesAdams2016,parmeeCooper2018,sutherlandRobicheaux2016,robicheauxSuresh2021,glicensteinBrowaeys2020,guerinKaiser2016} are inadequate to describe large arrays of quantum emitters -- even at a qualitative level, except at extremely-weak driving strengths.
In particular, in the thermodynamics limit of $N \rw \infty$, the threshold vanishes $\Omega \rw 0$, and there is no linear regime left at weak driving strength.
These conclusions motivate, and are supported by, the non-perturbative analysis discussed in the following.

\textit{A non-perturbative analysis.} 
To capture the steady state in the nonlinear (many-body) weak-drive regime discussed, we use the non-perturbative Dynamical Mean-Field Theory (DMFT) for Markovian spin systems introduced in \cite{scarlatellaCooper2024,scarlatellaSchiro2021}, for homogeneous steady states and in the thermodynamic limit $N\rw \infty$.
This is a bosonic version of DMFT, that maps the lattice model \eqref{eq:atLightME} onto an effective model of a single lattice site, coupled to an effective classical field and to a non-Markovian environment (see SM Note \ref{sm:dmft_formalism}), which are self-consistently determined, where the former is the analogous field of a Gutzwiller mean-field theory and the latter captures correlations beyond it.
DMFT becomes exact in the noninteracting limit (as well as in the limit of independent emitters; see SM Note \ref{sm:dmft_exact_limits}), which is most relevant in the weak-drive regime (at the same time, we recall that here the $\Omega \rw 0$ limit is non-perturbative, since we operate at $N\rw \infty$).

To solve the effective model, we use a method based on an expansion in the effective environment and a resummation of non-crossing diagrams (NCA), in a formulation that allows both a Markovian and a non-Markovian component \cite{schiroScarlatella2019,scarlatellaSchiro2021,scarlatellaSchiro2024a}. 
This is controlled when the effect of neighbouring sites in the effective single-site problem is small, which eventually limits our approach to a maximum interaction strength (a minimum value of $k_0 a$) above (below) which it does not converge. 
In the following we will set $k_0 a = 2$.

We found that solving the DMFT equations with a simple fixed-point iteration scheme did not converge at weak drive intensities, for any $k_0 a$ in the subwavelength regime. In this respect, this scheme fails even in the simpler case of a Gutzwiller mean-field theory, in the regime in which this predicts two simultaneous steady-states \cite{parmeeCooper2018} -- though this corresponds to smaller $k_0 a$s than considered here --, as one of them corresponds to a repulsive fixed point. To circumvent similar convergence problems, we implemented gradient-based methods (and/or a linear mixing scheme), which can in principle reach convergence at all fixed points (see SM Note \ref{sm:numerical_schemes}).

Using DMFT/NCA,  we predict a steady state with a finite density of subradiant excitation pairs.  This is shown in Fig. \ref{fig:fig1}~(c), reporting the number of steady-state excitations resolved in momentum, $n_{c}(k) = \aver{\sigma_k^+ \sigma_k^-}_c$, where $\sigma^-_k =  \sum_{ j } \ee^{\ii k r_j} \sigma^-_j  /\sqrt{N}$, and for the driving scheme considered (Fig. \ref{fig:fig1}~(b)) and at drive strengths of $\Omega \lesssim \Gamma$. 
Here the subscript ``c'' indicates a connected average $\aver{A B}_c \equiv  \aver{A B}- \aver{A} \aver{B} $, which only affects $n_{c}(k)$ at $k=0$ and will be discussed later.
Comparing with the overlaid dispersion relation, one sees that when the subradiant modes are resonant -- for $\Delta \gtrsim V_{k_0} \approx 0.1 \Gamma$ -- these are selectively populated. 
By contrast, when the radiative modes are resonant these do not get significantly populated as their finite lifetimes prevents the parametric instability to occur: instead, the population concentrates at the boundary of the subradiant manifold, corresponding to the modes closest to resonance yet with the smallest decay rates.
The linear stability of the steady-state was checked by enforcing a criterion for non-homogeneous fluctuations to remain bounded (see SM Note \ref{sm:dmft_linear_stability}).
We remark that the subradiant steady-state population has a finite density, of order of $n_{\rm c, subrad} = {\sum}_{{|k|>k_0}}\left\langle\sigma_k^{+} \sigma_{k}^{-}\right\rangle_c/N \approx 0.05$ for $\Omega /\Gamma \approx 1$ and $\Delta/\Gamma=0.4$, thus it is far from the zero-density, single-excitation regime \cite{asenjo-garciaChang2017}. 
We also recall that such population is solely bound by nonlinearities -- as it diverges in the effective linear model obtained perturbatively
-- resulting in interactions among its constituent excitations. 
Yet, similarly as in the single-particle regime \cite{asenjo-garciaChang2017}, this population has a well defined energy and momentum strongly-mismatched from free photons, thus it is subradiant.

\begin{figure*}
\centering
\includegraphics[width=0.75\linewidth]{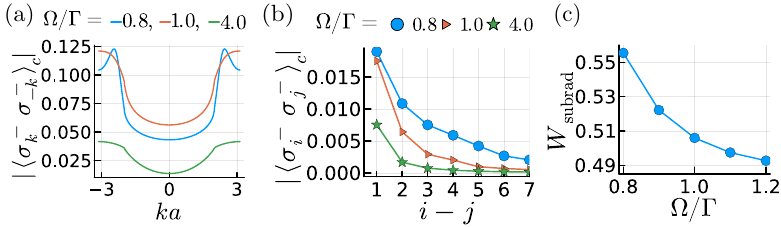}
\caption{{Pair-correlations function of the steady state computed in DMFT/NCA, for the driving conditions illustrated in Fig.~\ref{fig:fig1} and with a drive detuning of $\Delta/\Gamma=0.4$.
The subscript ``c"(``d'') indicates the connected (disconnected) component of such function, describing its nonlinear (linear) component. 
(a) The connected function as a function of momentum $k$ for different drive strengths $\Omega$: it shows multimode squeezing correlations of mode-pairs with momenta $(-k,k)$. Decreasing the drive strength it becomes more sharply peaked around the resonant modes.
(b) The non-local component of the connected function in real space, where $i-j$ is the lattice-sites distance, showing long-range correlations.
(a) and (b) also show that correlations increase in magnitude as the drive is decreased, as a non-trivial interplay of reduced occupations -- bounding correlations --, and reduced heating.  
(c) The ratio $W_{\rm subrad}$ (with a minus sign) between the connected component summed over the subradiant modes
$c_{c,\rm subrad} = \underset{{|k|>k_0}}{\sum}\left\langle\sigma_k^{-} \sigma_{-k}^{-}\right\rangle_c/N$ 
and the disconnected one 
$c_{\rm d} = \left\langle\sigma_{k=0}^{-}\right\rangle\left\langle\sigma_{k=0}^{-}\right\rangle/N$, showing that the nonlinear contribution to the steady-state is far from a small perturbation.
} 
}
    \label{fig:lowDrive}
\end{figure*}

In addition to populating the subradiant manifold, the nonlinear processes involved create excitations in correlated pairs. 
Nevertheless, such correlations could in principle be suppressed by heating due to interactions, driving and coupling with the vacuum environment. 
To assess this, we computed the connected pair-correlations function $\aver{\sigma^-_k \sigma^-_{-k}}_c$. 
Such function and its real-space counterpart $\aver{\sigma^-_i \sigma^-_{j}}_c$ are shown in panels (a) and (b) of Fig.~\ref{fig:lowDrive}. 
Panel (a) shows that correlations are indeed present in mode-pairs with momenta $-k$ and $k$, peaking around the resonant modes for decreasing drive strengths. 
These correspond to correlations that are long-ranged in real space, as shown in panel (b). 
Both panels also show a second important fact: the correlations amplitude increases decreasing the drive strength. This results from a non-trivial interplay between decreased populations and decreased heating. 
It is then clear that correlations can survive heating effects in such weak-drive regime.
Decreasing the drive further, we expect eventually a maximum and a trend inversion, as correlations become bound by occupations.
For those smaller drive strengths though, our method is not reliable and predicts negative probabilities due to an expected breakdown of the non-crossing approximation (NCA). 
For this reason, we limited our analysis to intermediate drive intensities for which our approach is well behaved.
We also remark that, although these correlations are reminiscent of a multimode bosonic squeezed state, the steady-state predicted is non-gaussian.
A qualitatively similar behaviour was also found for the correlator $\aver{\sigma^+_k \sigma^-_{k}}_c$.

Moreover, we wish to quantify the strength of nonlinearities in the weak-drive regime, and we consider their effect on the steady-state.
This state in fact features also an uncorrelated linear component, due to the linear driving, although detuned, of the zero-momentum mode, resulting in a coherent population witnessed by a uniform $\aver{\sigma_{j}^-} \neq 0$. It is the only component that would be predicted by classical theories \cite{ruostekoskiJavanainen1997,javanainenJavanainen1999}.


We estimate the relative importance of the nonlinear and linear components 
taking the ratio $W(A) = -c_{\rm c,A}/c_{\rm d}$, between the connected pair-correlation function summed over a momentum region $A$, $c_{\rm c,A} =  \sum_{k\in A} \aver{\sigma^-_k \sigma^-_{-k}}_c  /N $, and the disconnected component, $c_{\rm d} = \left\langle\sigma_{k=0}^{-}\right\rangle\left\langle\sigma_{k=0}^{-}\right\rangle/N $, carried by a delta function contribution at $k=0$. 
We note that if $A$ is the entire Brillouin zone, then the numerator is rigidly related to the denominator by a sum rule $-c_{\rm c} = \aver{\sigma^-_j}\aver{\sigma^-_j} =  c_d$, and $W =1$ regardless of the state. 
This results from the property $\aver{\sigma^-_i \sigma^-_{i}} =0 $ and is respected in DMFT/NCA.  
This sum rule shows that $W$ defines a distribution of the nonlinear correlations in momentum space that is normalised, such that it depends on the drive strength but its overall amplitude doesn't, and is thus interesting to track as a function of drive strength. 
In particular, we define $W_{\rm subrad}$ taking $A$ equal to the subradiant manifold ($A =\{k: \abs{k}>k_0\}$), which compares its nonlinear weight with the linear component and which is plotted in Fig.~\ref{fig:lowDrive}~(c).
This figure shows that $W_{\rm subrad}\approx 1/2$, indicating a strongly nonlinear state.
This indicates a high conversion efficiency of drive photons into the nonlinearly populated modes relevant for nonlinear optics. 
In addition, the trend with drive strength shows that the state becomes the more nonlinear the weaker the drive.

Although here we focused on the steady state, we also remark that the linear, uncorrelated component of the steady-state is superradiant and would be quickly dissipated by switching the drive off, leaving behind a slowly-decaying correlated subradiant state. 
In addition, nonlinearities also play and important role in shaping the dynamical correlation functions -- rather than the steady-state density matrix. This is shown in SM Note \ref{sm:correlation_functions_relaxation}, where the two-point local correlations are shown to fail to relax altogether if nonlinearities are neglected, thus those in turn completely determine the relaxation behaviour.

{
\textit{Conclusions -- }
We demonstrated that large subwavelength arrays of quantum emitters display a non-perturbative and strongly-nonlinear response down to arbitrary-weak drive intensities.
This is surprising, because the weak-drive regime of these systems had been widely regarded as a linear one, well described by classical equations~\cite{ruostekoskiJavanainen1997,javanainenJavanainen1999,leeRuostekoski2016,williamsonRuostekoski2020,bettlesAdams2015,bettlesAdams2016,parmeeCooper2018,sutherlandRobicheaux2016,robicheauxSuresh2021,glicensteinBrowaeys2020,guerinKaiser2016}. 
This effect is due to the important contribution of higher-order, resonant processes involving pairs of single-particle subradiant modes, which has been overlooked in previous analyses.
It depicts a scenario in stark contrast with the typical one of nonlinear optics, in which thick media and large drive powers are needed. 

Using a Dynamical Mean-Field Theory (DMFT), we showed that such regime leads to a steady state of interacting pairs of subradiant excitations, characterised by correlations with multimode squeezing and long-range character. 

At a technical level, obtaining consistent predictions at weak driving is notable, as this has proven extremely challenging for theoretical methods~\cite{minkFleischhauer2023,kramerRitsch2015,rubies-bigordaYelin2023}. 
Our DMFT/NCA approach was successful, albeit with the limitations discussed.


Our predictions can be realised in multiple experimental platforms, which directly implement the studied model: 
semiconductor dipolar excitons \cite{lagoinDubin2024}, already probing deep-subwavelength physics, and ultracold atoms~\cite{ruiBloch2020,srakaewZeiher2023}, which will soon operate in suitable regimes.

Our results have significant potential for applications.
First, they reveal a simple, experimentally-viable route to populate the subradiant states of quantum emitter arrays.
These are relevant for several purposes, notably photon storage \cite{plankensteinerGenes2015,facchinettiRuostekoski2016,asenjo-garciaChang2017,needhamOlmos2019,ferioliBrowaeys2021,ballantineRuostekoski2021,rubies-bigordaYelin2022,cechOlmos2023} and quantum metrology \cite{ostermannGenes2013,facchinettiRuostekoski2018,quRey2019}. However, their excitation is challenging, motivating a range of proposals~\cite{zannerKirchmair2022,jenChen2016,facchinettiRuostekoski2016,ballantineRuostekoski2021,rubies-bigordaYelin2022,cechOlmos2023} that nevertheless require fine experimental control such as single-atom resolution~\cite{zannerKirchmair2022,jenChen2016,facchinettiRuostekoski2016,ballantineRuostekoski2021,rubies-bigordaYelin2022,cechOlmos2023}, which is difficult to achieve in subwavelength regimes---in addition to the proposals being restricted to single-particle manipulation. 
By contrast, the nonlinear mechanism here discovered only requires a far-field coherent drive, routinely available in experiments. 

Second, our results demonstrate that the weak-drive regime of subwavelength quantum emitter arrays is far from being the classical regime previously thought~\cite{ruostekoskiJavanainen1997,javanainenJavanainen1999,leeRuostekoski2016,williamsonRuostekoski2020,bettlesAdams2015,bettlesAdams2016,parmeeCooper2018,sutherlandRobicheaux2016,robicheauxSuresh2021,glicensteinBrowaeys2020,guerinKaiser2016}, and it needs to be reconsidered. This is particularly relevant as it combines strong nonlinearities with weak driving powers and atom-thin structures, thereby opening a new avenue for nonlinear quantum optics \cite{krasnokAlu2018,liZentgraf2017,
solntsevKivshar2021,wildLukin2018}: 
by avoiding the significant heating inherent to conventional strong-driving regimes, this approach enables substantially stronger quantum correlations---as evidenced by the predicted correlated steady state---and supports far richer nonlinear quantum behaviour.

In this respect, the one-dimensional setup considered here may enable new avenues for entangled-photon generation~\cite{krasnokAlu2018,liZentgraf2017,solntsevKivshar2021,wildLukin2018}: the correlated pairs of subradiant atomic excitations can scatter into outgoing photons at the boundaries of a finite chain, where subradiant modes---despite being overall dark---couple to propagating photons~\cite{asenjo-garciaChang2017}. 
The outgoing photons can thereby be collected, potentially inheriting intricate multimode correlations from the atomic excitations. 

Finally, our prediction of a steady state exhibiting multimode squeezing correlations, combined with a clear and simple preparation scheme, is directly relevant to quantum metrology. Although such correlations are highly sought after~\cite{maNori2011}, viable preparation strategies remain scarce~\cite{kuriyattilDaley2025}. 
}

\textit{Acknowledgements -- } The Authors thank Mike Gunn for the stimulating discussions. 
This work was supported by the Engineering and Physical Sciences Research Council and Science and Technology Facilities Council [grant numbers EP/W005484, EP/V062654/1 and EP/Y01510X/1], and a Simons Investigator Award [Grant No. 511029].
For the purpose of open access, the authors has applied a creative commons attribution (CC BY) licence to any author accepted manuscript version arising.

The data (code) to reproduce the results of the manuscript will be provided by the author under request.

\bibliography{references}

\clearpage
\appendix
\setcounter{figure}{0}
\renewcommand{\thefigure}{S\arabic{figure}} 
\renewcommand{\appendixname}{SM Note}

\onecolumngrid

\vspace{\columnsep}
\begin{center}
{\Large\bf Supplementary Material for:} 

\vspace{0.2 cm}
{\large\bf ``How Subradiance Enables Nonlinearity in Weakly Driven Quantum Arrays''}

\vspace{0.5 cm}
Orazio Scarlatella$^{1,2}$
and Nigel R. Cooper$^{1}$
\vspace{0.3 cm}

$^{1}$\textit{TCM Group, Cavendish Laboratory, University of Cambridge, Cambridge CB3 0HE, UK} \\ 
$^{2}$\textit{Sorbonne Université, CNRS, Institut des NanoSciences de Paris,  75005 Paris, France}

\end{center}
\vspace{\columnsep}
\twocolumngrid

This Supplementary Material is organized as follows. 
In SM Note \ref{sm:perturbative_calculations} we take nonlinearities into account perturbatively, and show that this leads to an instability of the steady state.
In SM Note~\ref{sm:dmft_formalism} we discuss the main equations of the DMFT approach, in SM Note~\ref{sm:numerical_schemes} the numerical schemes to solve them, and in SM Note~\ref{sm:dmft_exact_limits} the exact limits of our DMFT approach. 
In SM Note~\ref{sm:dmft_linear_stability} we discuss the linear stability of the DMFT steady-state solution.
Finally, in SM Note~\ref{sm:correlation_functions_relaxation} we discuss the effect of interactions on time-dependent correlation functions, rather than on the steady-state density matrix. 

\section{Weakly-nonlinear theory of the weak-drive regime}
\label{sm:perturbative_calculations}

Here we develop a weakly nonlinear theory for the steady-state, in which nonlinearities are taken into account perturbatively with respect to the noninteracting theory, assuming a weak drive strength $\Omega$ and a large array size $N$. 
This is to some extent  similar to the treatment of a weakly-interacting Bose gas \cite{bogoliubovBogoliubov1947}. 

Under a weak driving, the steady state will have a low density of excitations $a^\dagger_i a_i \ll 1$ around the vacuum state with all spins pointing down. 
Then, we can take advantage of a Holstein-Primakoff representation of the spin operators around the lowest-weight state 
\begin{align*}
    \sigma_i^{-} &=\left(2 S-a_i^{\dagger} a_i\right)^{1 / 2} a_i, \\
    \sigma_i^{+} &=a_i^{\dagger}\left(2 S-a_i^{\dagger} a_i\right)^{1 / 2},   
    \\ \sigma_i^z &=2(a_i^{\dagger} a_i - S),
\end{align*}
and Taylor expand the square root with $S=1/2$ in our case. 
Keeping the lowest-order terms in the density, 
this leads to the Hamiltonian 
\begin{equation}
    \es{
    H &=  - \frac{ \Delta}{2} \sum_{i} {\sigma}_i^z +\frac{ \Omega}{2} \sum_{i} {\sigma}_i^x + \sum_{i  \neq j}   V_{i j} \sigma_i^{+} \sigma_j^{-}  \\ 
    &\approx \sum_{i} \Delta (\oh - a^\dagger_i a_i )+ \sum_{i \neq j}   V_{i j} a_i^\dagger a_j + \sum_{i} \frac{\Omega}{2} (a_i +  a_i^\dagger )   \\ 
     &\quad- \sum_{i} \frac{\Omega}{4} (  {a_i^\dagger a_i^\dagger a_i} + \hc ) - \sum_{i \neq j}   \frac{ V_{i j} }{2} (a_i^\dagger a_i^\dagger a_i a_j + \hc). 
    }
\end{equation}
In a momentum space representation, $a_{\mathbf{k}}=\sum_j \mathrm{e}^{\mathbf{i} \mathbf{k} \cdot \mathbf{r}_j} a_j / \sqrt{N}$ with $[a_{\mathbf{k}},a_{\mathbf{k}'}^\dagger] = \delta_{{\mathbf{k}},\mathbf{k}'}$, this takes the form: 
\begin{equation}
\label{sm_eq:bosonic_model}
    \es{
    H = &\sum_{k}  (-\Delta+V_k) a^\dagger_k a_k +  \frac{\sqrt{N}\Omega}{2} (a_{k=0} +  a_{k=0}^\dagger ) \\ 
    &- \sum_{k q} \frac{ \Omega}{4 \sqrt{N} } ( a_{k-q}^\dagger a_{q}^\dagger a_k + \hc ) 
    \\ &- \sum_{k q p}   \frac{ V_{k} }{2 N} (a_{k+q-p}^\dagger a_p^\dagger a_q a_k + \hc).
    } 
\end{equation}

Regarding both drive and nonlinear terms as perturbations in a time-dependent perturbation theory, populating the modes with $V_{\pm {k}} \neq \Delta $ involves non-resonant (i.e. energy non-conserving) processes that are energetically suppressed. 
We therefore retain only the resonant modes with $k = k_{\rm res}$ -- and drop the ``res'' label hereafter --  such that the problem reduces to a three-modes problem (including the driven $k=0$ mode). 

{We then use this model as the basis for a perturbative analysis, with respect to the noninteracting model discussed in the main text -- the first line of \eqref{sm_eq:bosonic_model}. 
First, we notice that the fluctuations of the $k=0$ mode around its average can be safely neglected.
Indeed, in absence of nonlinearities, the steady-state for the $k=0$ mode is a coherent state with
\begin{equation}
{\alpha}_0 = \aver{a_{k=0}} = {- \sqrt{N} \Omega }/[{ (-\Delta+V_{k=0})-\ii \Gamma_{k=0} ] =  \sqrt{N} \Omega \tilde{\alpha}_0},     
\end{equation}
which is proportional to $\sqrt{N} $ due to the direct coupling with the drive. We further assume $\sqrt{N} \Omega \ll \Gamma_{k=0} $ such that $a_{k=0}$ is small, and terms containing powers larger than two in this operator can be neglected.
Then, the fluctuations of $a_{k=0}$ in the steady-state are only determined by its (nonlinear) coupling with the other modes -- as they vanish in the coherent noninteracting steady state. 
Provided that these fluctuations are negligible with respect to the (macroscopic) average value $\alpha_0\sim \Omega \sqrt{N}$ we can replace the operator $a_{k=0}$ with its average. 
A posteriori, the 
nonlinear terms $\frac{\Omega}{4\sqrt{N}} a_{0} a^\dagger_k a^\dagger_{-k} $ and $ \frac{V_k}{2 N} a_0 a_{0} a_{k}^{\dagger} a_{-k}^{\dagger}$, 
which would lead to fluctuations of $a_0$, can also be neglected if $a_{\pm k} \lesssim \sqrt{N}$, as this ensures that they are small compared to the noninteracting Hamiltonian of the $k=0$ mode, of order $N \Omega^2$ -- this drive-independent threshold can always be enforced at a sufficiently weak driving, as $a_{\pm k}$ must decrease with the drive strength.
The substitution $a_{k=0} \rw \alpha_0$ then leads to an effective nonlinear model for the $a_{\pm k}$ modes.


To proceed with a perturbative analysis, one must be able to treat the further nonlinear terms in the operators $a_{\pm k }$ perturbatively, with respect to a noninteracting component of the model. Such a noninteracting part here is 
\begin{equation}
    \begin{aligned} 
H_k & \approx - \left(a_k^\dagger a_k + a_{-k}^{\dagger} a_{-k}\right)  \Omega^2 \left(  \re \tilde{\alpha}_0 +2 | \tilde{\alpha}_0|^2 V_0 +2 | \tilde{\alpha}_0|^2 V_k \right) 
\\ &  -  \left[ a_{-k}^{\da} a_k^{\da} \; \Omega^2 \tilde{\alpha}_0^2   
\left(\frac{1}{2} + {V_0}  + {V_k} \right)+\hc\right],
\end{aligned}
\end{equation}
where all contributions are of leading order $\Omega^2$, and independent of $N$.  
The same treatment is applied to obtain the dissipator. 
For a drive resonant with the subradiant modes, the decay rate of these modes is already small for large arrays -- $\Gamma \sim N^{-\alpha}$ ($\Gamma \sim \exp(-N)$) for open (closed) boundary conditions --, thus one can simply replace $\sigma_k^- \rw a_k$ in Eq. \eqref{eq:dissipat}, neglecting further terms from the Holstein-Primakoff expansion.

The noninteracting part of the effective model is then that of a two-modes parametric amplifier~\cite{scullyZubairy1997}, with Hamiltonian $H = \delta (a^\da a + b^\da b) + {\lambda} ( a^\da b^\da +  \hc ) $ -- where we dropped the $k$ indices.
This is well known~\cite{scullyZubairy1997} to undergo a parametric instability for a drive strength $|\lambda|^2 > \delta^2 + \Gamma^2$, accompanied by a divergence of steady-state moments (its populations and anomalous correlators). 
In our case,
this instability occurs for $\Omega \gtrsim N^{-\alpha/2}$ ($\Omega \gtrsim \exp(-N/2)$) for open (closed) boundary conditions.
For the most subradiant modes, the exponent $\alpha$ was found to be $\alpha \approx 3$ in 1D, and $\alpha \approx 6$ or $3$ in 2D \cite{asenjo-garciaChang2017}. 

Beyond this threshold, the noninteracting
model is dynamically unstable, and it therefore cannot be
used as the basis of a perturbative analysis. The steady-state must then be obtained using non-perturbative approaches.
}

\section{Nonequilibrium steady-state DMFT}
\label{sm:dmft_formalism}


In the present case, the effective single-site model obtained in DMFT (impurity model) is a generalized spin-boson model, which in the steady state is described by the time-translation invariant Keldysh action 

\begin{equation}
\es{
\label{eq:n_sbAction}
S^{\mathrm{imp}}&=S_{0}^{i}+\oh  \int_{-\infty}^\infty d t \lsq \varsigma_{i}^\da(t) \tau_3 b  + \hc \rsq \\ 
&-\frac{1}{2} \int_{-\infty}^\infty d t \int_{-\infty}^\infty d t^{\prime} \varsigma_{i}^\da(t)  \tau_3 {\mathcal{W}}\left(t-t^{\prime}\right) \tau_3 \varsigma_{i}\left(t^{\prime}\right) 
}
\end{equation}
Here $\varsigma_i = (d_{i+}, \bar{d}_{i+},d_{i-}, \bar{d}_{i-} )^T$ is a vector in Nambu and Keldysh formalisms, where $d=\aver{\sigma^-}$ and $\bar{d}=\aver{\sigma^+}$ are average values on spin coherent states \cite{altlandSimons2012}, and the $+$ and $-$ indices indicate to which branch of the Keldysh double contour the fields belong to.
$b$ and $\mathcal{W}$ represent the effective field and effective environment, which are site-independent for a homogeneous steady state.
$\tau_3$ is a diagonal matrix with entries $(1,1,-1,-1)$ arising in the Keldysh-Nambu formalism, where a Nambu formalism is needed to allow for nonzero values of the anomalous correlators $\left\langle\sigma^{ \pm} \sigma^{ \pm}\right\rangle$, which arise due to the drive term, breaking the $U(1)$ symmetry of the undriven problem. 

The self-consistent conditions correspond to a set of matrix relations in frequency and momentum space: 
\begin{align}
\label{eq:n_dmft1}
    \Pi_{\mathrm{loc}}&=\left[ \tau_3{\chi}^\mo_{ii}\tau_3  + \mathcal{W}\right]^{-1}  \\ 
    \label{eq:n_dmft2}
    U_{ii}&=\frac{1}{N} \sum_{k} \left[W_{k}^\mo -\Pi_{\mathrm{loc}} \right]^{-1} \\ 
    \label{eq:n_dmft3}
    \mathcal{W}^{-1}&=\Pi_{\mathrm{loc}}+U_{ii}^{-1} \\ 
    \label{eq:n_dmft4}
    b_{\pm}&=\left(  \mathcal{W}^R(\w={0})-  W_{k=0}^R(\w=0)\right) \aver{\varsigma_{i\pm}}.
\end{align}
Here the superscript ``$R$'' indicates the retarded component and $b_\pm$ and $\aver{\varsigma_{i\pm}}$ the $+$ or $-$ components of the corresponding vectors.  
$W_k$ is a known matrix with entries defined by the Fourier transforms of \eqref{eq:vij} and \eqref{eq:gamij}
\begin{equation}
\label{eq:wk}
   \tau_3 W_k \tau_3   =  \bma   {V}_{k} - \frac{\ii \Gamma_k }{2}  & 0 & 0 & 0 \\ 
0 &    {V}_{k} - \frac{\ii \Gamma_k }{2}  & 0 & i \Gamma_{k} \\ 
i  {\Gamma_{k}} & 0 & -   {V}_{k} - \frac{\ii \Gamma_k }{2}  & 0 \\
0 & 0 & 0 &  -  {V}_{k} - \frac{\ii \Gamma_k }{2} 
\ema
\end{equation}
Assuming one can compute from the impurity action \eqref{eq:n_sbAction} the local connected Green's function $
\chi_{ii}(\tau)= -\ii \lim_{t\rw \infty}\langle  \varsigma_{i}(t+\tau) \varsigma^\da_{i}(t)\rangle_{S_{\text {imp }}^{\text {con }}}$
and expectation value $\left\langle\varsigma_{i}\right\rangle$, then these equations determine $\mathcal{W},b,\Pi_{\mathrm{loc}}$ and $U_{ii}$, such that the impurity problem represents the original lattice problem. 

The connected propagator of the lattice $\chi_{ij}(\tau)= -\ii \lim_{t\rw \infty} \langle  \varsigma_{i}(t+\tau) \varsigma^\da_{j}(t)\rangle^{\textrm{con}}$ can then be computed  by $\chi^\mo(k,\omega) = \tau_3 (\Pi_{\rm loc}^\mo(\omega) - W_k)\tau_3$.

We notice that this DMFT approach reduces to Gutzwiller mean-field theory fixing $\mathcal{W}(\w) = 1/N \sum_k W_k = W_{ii} $.

\section{Schemes for solving the DMFT equations}
\label{sm:numerical_schemes}

The usual procedure to solve the DMFT self-consistent equations is by a fixed-point iteration scheme. Here we go beyond this scheme and implement a linear-mixing and a gradient-based Broyden method. 
In the following, we first describe the fixed-point iteration scheme. \\

\subsection*{Fixed-point iteration scheme}
Starting from a guess for $b$ and  $\mathcal{W}$, such as the mean field solution for the first and the single-atom decay for the second $\mathcal{W} = 1/N \sum_k W_{k}$, the following steps are iterated until a fixed point is reached: 

\begin{itemize}
\item \textit{Solving the impurity model}: given $\mathcal{W}$, $b$, the spin model is solved using the impurity solver, computing the steady state density matrix $\rho_s$ and atomic correlation function 
$\chi = \bma \chi^{{++}}& \chi^{+-} \\ \chi^{-+} & \chi^{--}  \ema$ 
for $t>0$, where the entries are $2\times 2$ matrices in Nambu space. 
Then $\chi$ at negative times $t<0$ is obtained assuming the steady state relation 
\begin{equation} 
\label{eq:ssCorrFuncs}
\chi_{\alpha \beta}^{ab}(-t) = - [\chi^{ba}_{\bar{\beta}\bar{\alpha}}]^* (t)
\end{equation}
where the conjugate-transpose of both Nambu $a,b$ and Keldysh $\alpha \beta$ indices is taken and the Keldysh indices $\alpha,\beta \in [+,-]$ are negated, such that $\bar{\alpha} = -\alpha$.
\item \textit{updating the Weiss field $\mathcal{W}$ and effective field $b$}: we evaluate the self-consistent equations \eqref{eq:n_dmft1},\eqref{eq:n_dmft2},\eqref{eq:n_dmft3} and \eqref{eq:n_dmft4} where in frequency and momentum space inverses of Green's functions are simply given by a matrix inverse of their $4 \times 4$ Nambu-Keldysh stucture. These give new values for $b$ and $\mathcal{W}$, whichare transformed back into real time to interate the procedure. 

\end{itemize}

An important point of the update procedure concerns how to define the Green's functions at time $t=0$ such as to interface the NCA impurity solver and Keldsyh field theory. This is discussed in \cite{scarlatellaCooper2024}. \\ 

\subsection*{Linear-mixing and Broyden schemes}
One DMFT step in a fixed-point iteration scheme can be considered as a functional $G$ of the input hybridization function $\mathcal{W}$ and effective field $b$, i.e.
$
(\mathcal{W}^{\text {new }},b^{\text {new }})= G \left\{\mathcal{W}^{\text {old}},b^{\text {old}}\right\}$. 
This is iterated until a fixed point is reached. 
Defining a mapping $F$ as the difference
$F\lbr \mathcal{W},b\rbr= G \left\{\mathcal{W},b\right\} - (\mathcal{W},b)$
the approach to self-consistency clearly corresponds to solving the system of equations
$
F\lbr \mathcal{W},b\rbr=0
$. 
A fixed point iteration scheme converges if all the eigenvalues of the linearization of $G$ in the vicinity of the fixed point are stictly less than $1$ in absolute value \cite{zitkoZitko2009}. 
This might not be the case for all solutions of the DMFT equations $F$ and a fixed-point iteration scheme might not converge despite these being physical solutions \cite{zitkoZitko2009,ganahlVerstraete2015}.
In such cases, convergence can be achieved in some cases by a linear mixing scheme~\cite{zitkoZitko2009}, or by gradient-based methods that guarantee convergence to all solutions~\cite{zitkoZitko2009}. 

First let's consider the equation for $b$. This reduces to the equation of a Gutzwiller mean-field theory if we keep $\mathcal{W} = \ 1/N \sum_k W_k$ fixed. In such case, we found that a fixed-point iteration does not converge to all solutions when Gutzwiller mean-field predicts two simultaneous steady-states \cite{parmeeCooper2018}, as one of them corresponds to an unstable fixed point. 
To solve this problem, we implemented a Newton method to determine $b$, where luckily the gradient can be computed analytically.
The problem corresponds to finding the zero of the function $\tilde{F}(b) = b - \left(\mathcal{W}^R(\omega=0)-W_{k=0}^R(\omega=0)\right)\langle\varsigma\rangle$ (\eqref{eq:n_dmft4}). In a Newton method, the update is $b_{\rm new} = b_{\rm old} - J^\mo(b_{\rm old}) \tilde{F}(b_{\rm old})$, where $J$ is the Jacobian of $\tilde{F}$. To compute the Jacobian, we use the Kubo formula $\langle\varsigma\rangle = - \chi^R(\w=0) \delta b$ relating the steady-state average values and the perturbation $\delta b$ (the minus sign reflect that of $b$ in the Hamiltonian), leading to $J = \id + \left(\mathcal{W}^R(\omega=0)-W_{k=0}^R(\omega=0)\right) \chi^R(\w=0)$ (note that non-connected Green functions should be used instead of connected, but for the retarded component in the steady state these are the same).

Even with this method, we found that convergence of DMFT slows down decreasing the drive strength in the subwavelength regime of $k_0 a < \pi$, and below a certain value no solution is found. 
For this reason, we went beyond a fixed point scheme also to determine the hybridization function $\mathcal{W}(\omega)$. 
A simple improvement is achieved by mixing the previous and new guesses for $\mathcal{W}$, indexed as $m$ and $m-1$, in the fixed-point iteration scheme: 
$$
\mathcal{W}^{\text {input, }(m)}=\alpha \mathcal{W}^{\text {new,(m) }}+(1-\alpha) \mathcal{W}^{\text {input,(m-1) }}
$$
where $\alpha \in [0,1]$ is a mixing parameter. 
This method allowed to reach smaller drive strengths, yet requiring an increasingly-smaller $\alpha$ the smaller the drive strength, and could not converge below a certain point. 
Indeed, a linear mixing approach still does not guarantee convergence to all solutions \cite{zitkoZitko2009}.

To rule out such a problem we also implemented a Broyden's method, following \cite{zitkoZitko2009}. 
This estimates the gradient in an iterative manner, as this cannot be computed analytically in the case of $\mathcal{W}$. 
This method reduces to linear mixing for a gradient proportional to identity $J = -\id /\alpha$, where $\alpha$ is the analogous of a linear mixing parameter. There are other arbitrary parameters in the method, that we kept fixed as in~\cite{zitkoZitko2009}.

An example of convergence at weak-drives is shown in Fig. \ref{fig:broy}, where the DMFT error at the $m$-th iteration is defined as 
\begin{equation}
\es{
e(m) = &{\textrm{max}_{t_i,\eta,\zeta}}( \abs{[\mathcal{W}_m(t_i)]^{\eta \zeta} - [\mathcal{W}_{m-1}(t_i)]^{\eta \zeta}} ) \\ &+ {\textrm{max}_{\eta}}{\lp b_m^\eta - b_{m-1}^\eta \rp}
}    
\end{equation}
where $\eta,\zeta$ are Keldysh-Nambu indices. 
One sees that Broyden's algorithm speeds up convergence with respect to linear mixing. 
We note that the error saturates to a small value instead of decreasing monotonically, and this is due to finite numerical accuracy (mainly timestep size and momentum-integration errors), but those results can be considered converged.
One also sees that a small mixing parameter $\alpha$ is needed for convergence in the linear mixing case, resulting in a slow convergence, while a larger one can be used with Broyden's algorithm.

\begin{figure}
    \centering
    \includegraphics[width=1\linewidth]{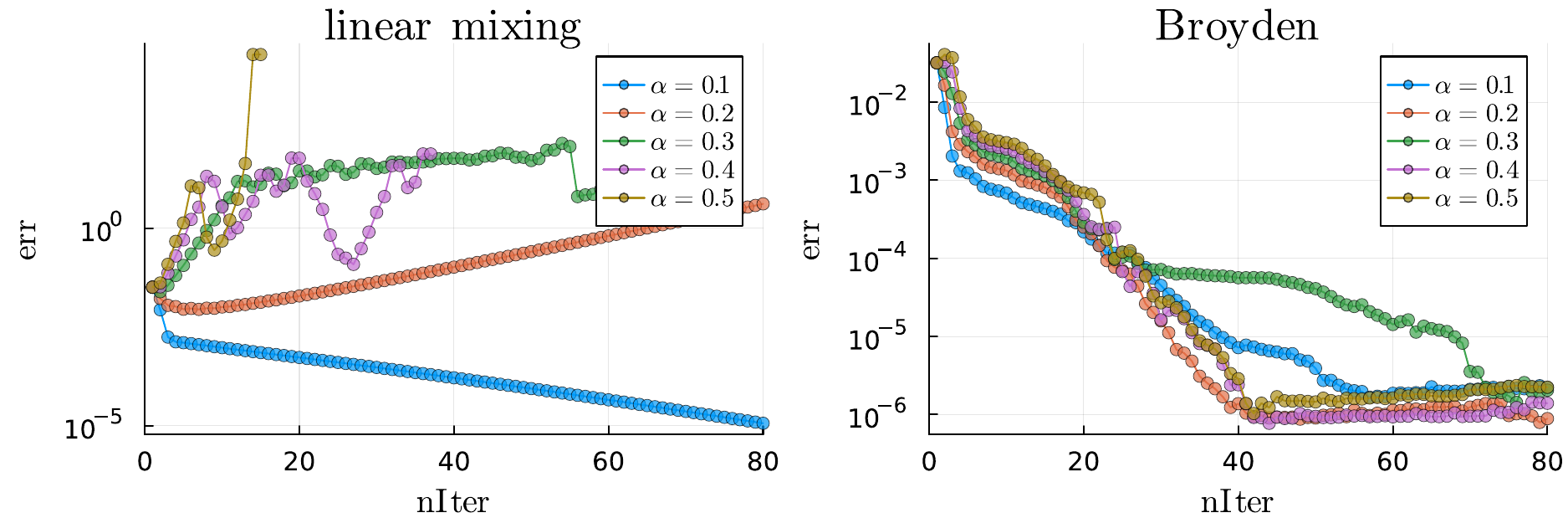}
 \caption{DMFT error as a function of number of DMFT iterations, starting close to the previously found solution at $k_0=2$ and $\Omega=0.3, \Delta=0$. Here $dt=0.1$ and $t_{\rm max}=200$. }
    \label{fig:broy}
\end{figure}
Eventually, Broyden's method did not allow to converge in a significantly broader parameter regime than a simpler linear-mixing scheme, namely 
at lower drive intensities and smaller $k_0 a$. 
Convergence is in fact eventually limited by the non-crossing approximation NCA that starts breaking down predicting negative probabilities as those regimes are approached, which is expected as these are highly collective regimes far away from independent emitters behaviour.

\section{Exact limits of DMFT}
\label{sm:dmft_exact_limits}

Here we discuss two exact limits of DMFT, in the context of our formulation. 
The first is the usual atomic limit -- here the limit of independent emitters. In this limit the impurity problem becomes equivalent to a single site of the original lattice problem: the impurity action \eqref{eq:n_sbAction} reduces to a single-site action with $b=0$ and $\mathcal{W} = 1/N \sum_k W_{k}$ describing the local dissipation. In addition, in this limit the NCA solver we use \cite{scarlatellaCooper2024,scarlatellaSchiro2021,schiroScarlatella2019} exactly reduces to a Lindblad master equation, introducing no further approximations on the solution of the single-site problem.

In addition, DMFT becomes exact at low-densities, where the model reduces to a model of noninteracting bosons.
To show this, we refer to the derivation of the approach, reported in \cite{scarlatellaCooper2024}, or in \cite{lenkEckstein2022a}. DMFT is formulated on a lattice problem equivalent to Eq.  \eqref{eq:atLightME}, in which auxiliary bosonic degrees of freedom are introduced by decoupling the non-local terms (the $V_{ij}$ and $\Gamma_{ij}$ terms) with a Hubbard-Stratonovich transformation.
The resulting boson-spin lattice problem action reads 
$S_{\mathrm{HS}}=S_{0}+S_{\phi \phi}+S_{\phi \sigma}$, where $S_{0} = \sum_r S_{0,r}$ is the action of decoupled spins at position $r$, 
$S_{\phi \phi}$ is a free-bosons action in terms of a Nambu vector of complex bosonic fields $\phi_r$
$$
S_{\phi \phi}= \oh \int_{-\infty}^\infty  d t  d t^{\prime} \sum_{r r^{\prime}} \phi_{r}^\da(t)\tau_3 \left[W^{-1}\right]_{r r^{\prime}}\left(t-t^{\prime}\right) \tau_3 \phi_{r^{\prime}}\left(t^{\prime}\right),
$$
and  $W_{rr'}(t-t')$ is the Fourier transform of \eqref{eq:wk}. $S_{\phi \sigma}$ describes a \textit{local} linear coupling term between bosons and spins (with spin Nambu fields $\varsigma$ defined as in \eqref{eq:n_sbAction})
$$
S_{\phi \sigma}= \oh \sum_{r} \int_{-\infty}^\infty d t \phi^\da_{r}(t) \tau_3 \varsigma_{r}(t) + \hc
$$
In the case in which nonlinearities can be discarded, the spin variables can be replaced by bosonic ones, resulting into a gaussian action of a single atom that can be integrated out exactly. Importantly, this results into a self-energy for the auxiliary bosons that is local in space $\Pi_{r r'} = \Pi_{\rm loc} \delta_{r r'}$-- this is the case because the atoms are only coupled to each other through the auxiliary bosons. 
One can also show that the boson-spin lattice model, being entirely quadratic, can be exactly mapped onto the impurity model \eqref{eq:n_sbAction} with a cavity construction where all other sites are integrated out. Since these are the two assumptions on which DMFT rests \cite{georgesRozenberg1996,aokiWerner2014}, this approach becomes exact in this noninteracting regime. 
This is shown numerically in Fig. \ref{fig:keldysh_k0a=pi}, where the local Green's functions computed via DMFT and the noninteracting bosons model agree, outside of the subwavelength regime (for $k_0 a > \pi$). 
We remark that the non-crossing approximation (NCA) of the impurity problem introduces a further approximation, but here this is controlled because the interactions between different emitters are relatively weak. The two quantities instead are shown to deviate in the subwavelength regime where interactions are stronger ($k_0 a < \pi$) in Fig. \ref{fig:keldysh}.
This eventually limits our DMFT/NCA approach at low $k_0 a$ values and weak drives, for which the emitters behave strongly collectively. 
 
\begin{figure}
    \centering
    \includegraphics[width=0.65\linewidth]{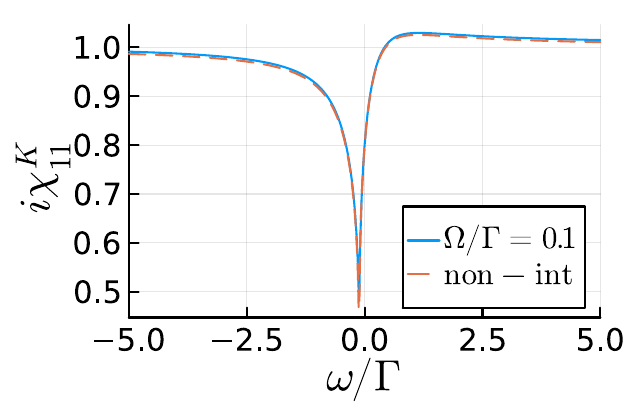}
    \caption{The local atomic Keldysh Green's function defined as in Fig. \ref{fig:keldysh} computed from a noninteracting theory for $\Omega\rightarrow 0$ (dashed line) and in DMFT (solid), for $k_0a =\pi$ and $\Delta=0$. This deviates from a Lorentzian shape for a single atom due the dipolar interactions between the emitters.}
     \label{fig:keldysh_k0a=pi}
\end{figure}

\section{Linear stability equations}
\label{sm:dmft_linear_stability}

The homogeneous DMFT equation \eqref{eq:n_dmft4} can be generalized to allow a momentum and frequency dependent effective field, allowing to study the linear stability of such perturbations. 
Note that, in order to obtain a condition that can be evaluated from the homogeneous solution only, we don't also allow for an inhomogeneous perturbation of the effective bath, that we keep instead homogeneous.

The field must then satisfy the self-consistent condition $b_k(\w) = \lsq \mathcal{W}^R(\w) -W_k^R(\w) \rsq  \aver{\varsigma_{k} \left( \w \right)}$ and, assuming a perturbation around a homogeneous value  $b_k(\w) = b \delta(k) \delta(\omega) + \delta b_{k}(\w)$, the steady-state expectation $ \aver{ \varsigma_k(\w) } = \aver{\varsigma } \delta(k) \delta(\omega) + \aver{\delta \varsigma_k(\w) } $ can be obtained from linear response $\aver{\delta \varsigma_k(\w) } = - \chi_k^R(\w) \delta b_{k}(\w) $.
Here ``$R$'' indicates the retarded components of $\mathcal{W}$, representing the effective environment, $\chi$ is the local atomic Green's function and $W_k$ a matrix of the coherent and dissipative couplings \eqref{eq:vij} and \eqref{eq:gamij} defined in Eq.~\eqref{eq:wk}. 
Then, we obtain that a finite $\delta b_{k}(\w)$ can form if the following condition is satisfied: 
\begin{equation}
\label{eq:dmft_stab}
\det \lbr \lsq W^R_k - \mathcal{W}^R(\w) \rsq^\mo -\chi^R(\w) \rbr = 0.
\end{equation}
A similar equation determines the linear stability of a Gutzwiller mean-field theory by fixing $\mathcal{W}(\w) = 1/N \sum_k W_k $. 
Using these equations, we checked that the steady-state solution predicted in DMFT is stable to perturbations. 
We also note that, while a Gutzwiller mean-field approximation predicts instabilities of the uniform steady-state solutions towards several non-uniform phases \cite{parmeeCooper2018}, these only occur at smaller values of  $k_0 a \lesssim 1$ than those considered here. For those values and in the weak-drive regime, our DMFT/NCA approach does not converge, thus the fate of those instabilities and phases upon including correlations remains an interesting open problem.

\section{Relaxation of local correlation functions} 
\label{sm:correlation_functions_relaxation}

In the simplest noninteracting bosons approximation, the steady-state single-particle Green's functions are easily obtained in a Keldysh formalism. Assuming the thermodynamic limit $N\rw \infty$, in the limit of vanishing drive $\Omega \rw 0$ the problem becomes diagonal in Nambu space and one gets the inverse Green's function in the basis of classical and quantum fields
\begin{equation}
\label{eq:lowOmGreen}
\chi_{11}^\mo(k,\w) =  \bma 0 & \w  + {\Delta} -{V}_{k} - \frac{\ii}{2} {\Gamma_{k}} \\
\w + {\Delta}- {V}_{k} + \frac{\ii}{2} {\Gamma_{k}} & \ii \Gamma_{k}  
\ema  
\end{equation}

In the presence of subradiant modes with zero damping, in the subwavelength regime, whether local correlation functions can relax to a steady state is a non-trivial question. 
These probe the relaxation of a local excitation created in the steady state, which therefore is spread across all momentum modes and whose long-time dynamics is dominated by the non-radiative modes. 
Local correlation functions are obtained by a matrix inversion of \eqref{eq:lowOmGreen} and a sum over momentum $\chi_{11} = \sum_k \chi_{11}(k,\w)$. We consider for example the Keldysh component, corresponding to the top-left entry of the matrix, defined as $\ii \chi_{11}^K(\tau) =  \lim_{t\rightarrow\infty}  \aver{ \left\lbrace \sigma_i^-(t+\tau),\sigma_i^+(t) \right\rbrace } - \left\lbrace \aver{ \sigma_i^-(t+\tau)},\aver{ \sigma_i^+(t) } \right\rbrace $. 

Even in a noninteracting theory, where the decay rate of non-radiative modes is zero, such correlation function could in principle still relax to a steady state by dephasing of the different modes, constituting a continuum. 
In Fig. \ref{fig:keldysh} we show that nevertheless these fail to relax.
An interesting question then is whether taking nonlinearities into account, which can provide an effective lifetime for the non-radiative modes, allows these quantities to relax. 

In DMFT/NCA we find that these quantities do relax and correspond to a smoothened version in frequency of those predicted by a noninteracting theory, as shown in Fig. \ref{fig:keldysh}. 
These predictions might be nevertheless limited by inaccuracies of the non-crossing approximation in this regime. 
The prediction of the noninteracting theory where an artificial damping is added was also useful in the present work to initialize the DMFT. 
\begin{figure}[h]
    \includegraphics[width=1.0\linewidth]{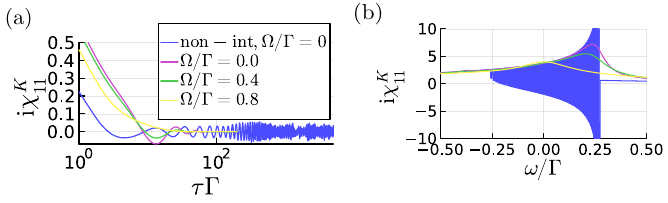}
    \caption{The local atomic Keldysh Green's function $\ii \chi_{11}^K(\tau) =  \lim_{t\rightarrow\infty}  \aver{ \left\lbrace \sigma^-(t+\tau),\sigma^+(t) \right\rbrace } - \left\lbrace \aver{ \sigma^-(t+\tau)},\aver{ \sigma^+(t) } \right\rbrace $ computed from a noninteracting theory for $\Omega\rightarrow 0$ and in DMFT in real time (a) and frequency (b), for the same parameters as in Fig. \ref{fig:lowDrive}~(c). 
    }
    \label{fig:keldysh}
\end{figure}

\end{document}